# Tellurium Hydrides at High Pressures: High-temperature Superconductors

Xin Zhong[1], Hui Wang[1], Jurong Zhang[1], Hanyu Liu[1], Shoutao Zhang[1], Hai-Feng Song[4,5], Guochun Yang[2,1,*], Lijun Zhang[3,1,*], Yanming Ma[1,*]

[1]*State Key Laboratory of Superhard Materials, Jilin University, Changchun 130012, China.*

[2]*Faculty of Chemistry, Northeast Normal University, Changchun 130024, China.*

[3]*College of Materials Science and Engineering and Key Laboratory of Automobile Materials of MOE, Jilin University, Changchun 130012, China*

[4]*LCP, Institute of Applied Physics and Computational Mathematics, Beijing 100088, China*

[5]*Software Center for High Performance Numerical Simulation, China Academy of Engineering Physics, Beijing 100088, China*

Observation of high-temperature superconductivity in sulfur hydrides at megabar pressures has generated an irresistible wave on searching for new superconductors in other compressed hydrogen-rich compounds. An immediate effort is towards exploration of the relevant candidate of tellurium hydrides, where tellurium is isoelectronic to sulfur but it has a heavier atomic mass and much weaker electronegativity. The hitherto unknown phase diagram of tellurium hydrides at high pressures was investigated by a first-principles swarm structure search. Four energetically stable and metallic stoichiometries of $H_4Te$, $H_5Te_2$, $HTe$ and $HTe_3$ were uncovered above 140 GPa, showing a distinct potential energy map of tellurium hydrides from those in sulfur and selenium hydrides. The two hydrogen-rich $H_4Te$ and $H_5Te_2$ species adopt ionic structures containing exotic quasi-molecular $H_2$ and linear $H_3$ units, respectively. Strong electron-phonon couplings associated with the intermediate-frequency H-derived wagging and bending modes make them good superconductors with high $T_c$ in the range of 46-104 K.

A recent breakthrough finding in the superconductivity field is the observation of remarkably high superconductivity (with $T_c$ up to 190 K) in sulfur dihydride ($H_2S$) under pressure [1]. This observation was achieved by a direct investigation on a theoretical prediction of high-$T_c$ superconductivity in compressed solid $H_2S$ within the framework of Bardeen–Cooper–Schrieffer (BCS) theory [2,3]. The superconductive mechanism of $H_2S$ and its possible decomposition at high pressures was then substantially explored [4-9]. Besides these efforts, findings of new superconductors in other relevant hydrogen-containing compounds have also attracted great attention. Selenium (Se) hydrides were already predicted to exhibit high $T_c$ in the range of 40-131 K at megabar pressures [10, 11].

Tellurium (Te) is the next group-VI element isoelectronic to S and Se. However, Te adopts an even larger atomic core with a much weaker electronegativity, and therefore it exhibits a rather different chemistry from S and Se. As a result, stable $H_2S$ [12] and $H_2Se$ [13] gas molecules and their solid counterparts exist at ambient pressure, whereas $H_2Te$ gas molecules are unstable and rapidly decompose into the constituent elements (above −2 °C) [14]. Thus far, there is lack of any report on stable Te hydrides.

Pressure can fundamentally modify chemical reactivity of elements, and overcome the reaction barrier of hydrogen and certain substances to form stable hydrides (*e.g.* noble metal hydrides [15,16], $LiH_6$ [17], $NaH_9$ [18], and $CaH_6$ [19], etc). There is a possibility that Te hydrides can be synthesized by compressing a mixture of Te + $H_2$. As to the superconductivity, on one hand one may argue that Te hydrides might not be good candidates for high-$T_c$ superconductors since the low Debye temperature caused by heavy Te can suppress the superconductivity. On the other hand, low-frequency vibrations (soft phonons) associated with a larger atomic mass can enhance electron-phonon coupling (EPC) [20] as seen from the predicted higher $T_c$ (up to 80 K) in $SnH_4$ [21] than those in $SiH_4$ (up to 17 K) [22] and $GeH_4$ (up to 64 K) [23].

We herein extensively explored the high-pressure phase diagram of Te hydrides by using the swarm-intelligence based CALYPSO structural prediction calculations [24,25]. Distinct from S and Se hydrides, Te hydrides exhibit a unique potential energy landscape, where the unexpected stoichiometries of $H_4Te$, $H_5T_2$ and $HTe_3$ emerge as stable species at megabar pressures. $H_4Te$ is so far the most H-rich stoichiometry reported in the family of chalcogen hydrides. The uncovered H-rich $H_4Te$ and $H_5T_2$ compounds consist of unforeseen quasi-molecular $H_2$ and linear $H_3$ units, respectively, both of which are metallic and show high-temperature superconductivity with $T_c$ reaching as high as 104 K at 170 GPa for $H_4Te$.

The in-house developed CALYPSO structure prediction method designed to search for the stable structures of given compounds [24,25] has been employed for the investigation of phase stability of Te hydrides at high pressures. The most significant feature of this methodology is its capability of unbiasedly predicting the ground-state structure with only the knowledge of chemical composition. The details of the search algorithm and its several applications have been described elsewhere [24,26-28]. The underlying energetic calculations were performed within the framework of density functional theory using the plane-wave pseudopotential method as implemented in the VASP code [29]. The Perdew–Burke–Ernzerhof generalized gradient approximation [30] was chosen for exchange-correlation functional. The electron-ion interaction was described by projected-augmented-wave potentials with the $1s^1$ and $5s^25p^4$ configurations treated as valence electrons for H and Te, respectively. During the structure search, an economy set of parameters were used to evaluate the relative enthalpies of sampled structures on the potential energy surface, following which the kinetic cutoff energy of 500 eV and Monkhorst−Pack k-meshes with grid spacing of $2\pi \times 0.03$ Å$^{-1}$ were chosen to ensure the enthalpy converged to better than 1 meV/atom. The phonon spectrum and EPC of stable compounds were calculated within the framework of linear response theory through the Quantum-ESPRESSO code [31].

We focused our structure search on the H-rich compounds promising for higher-$T_c$ superconductivity [32]. The energetic stabilities of a variety of $H_xTe_y$ ($x = 1 − 8$ and $y = 1 − 3$) compounds were evaluated through their formation enthalpies relative to the products of dissociation into constituent elements ($\Delta H$) at 0, 50, 100, 200, and 300 GPa, as depicted in Fig. 1 and Supplementary Fig. S1. At 0 GPa, no stoichiometry is stable against the elemental dissociation, consistent with the fact of non-existence of any solidified H-Te phase at ambient condition. This situation preserves up to 100 GPa, but accompanied with much decreased $\Delta H$ magnitudes, implying a tendency of H-Te compounds being stabilized under further compression. Indeed at 200 GPa, all the stoichiometries become energetically favored over the elemental dissociation, and stable stoichiometries of $H_4Te$, $H_5Te_2$ and HTe emerge on the convex hull. At 300 GPa, in addition to these three species, another stable stoichiometry of $HTe_3$ appears in the H-poor regime. The finding of these unexpected stable H-Te stoichiometries is in sharp distinction from the actual stabilization of $H_3S/Se$, HS/Se, $HS_2/Se_2$ in S and Se hydrides [9,10]. As will be discussed later, all the four stable species are metallic in their stable pressure regions (Fig. 1b).

The most H-rich $H_4Te$ was predicted to crystallize in a hexagonal structure (space group $P6/mmm$) above 162 GPa, consisting of H-sharing 12-fold $TeH_{12}$ octahedrons (Fig. 2a). The exotic structure feature of this phase is that the short H-H contact (~0.85 Å at 200 GPa) between two octahedrons forms the quasi-molecular $H_2$-unit. Analysis of the electron localization function (ELF) between two H atoms within the $H_2$ unit shows a high value (~0.9, see Fig. S3a), indicating a strong H-H covalent bond. The longer H-H bond length than that of free $H_2$ molecule (0.74 Å) is attributed to the accepted charge of ~0.44e per $H_2$ donated by Te (Table S3), which resides in the anitbonding orbital of $H_2$ and thus lengthens the intramolecular bond. Note that such charge transfer is a prerequisite on the formation of the quasi-molecular $H_2$ in compressed hydrides, as also reported in $LiH_n$ ($n = 2, 6,$ and 8) [17], $CaH_6$ [19], $GeH_4$ [23], and $SnH_4$ [21,33]. No electron localization is found between H and Te in the ELF calculations, signaling an ionic H-Te bonding.

Upon compression, the $P6/mmm$ structure of $H_4Te$ transforms into an energetically more favored $R-3m$ structure at 234 GPa. This transition is a first-order in nature and is accompanied by an increased coordination number of Te from 12 to 14 with the formation of H-sharing $TeH_{14}$ octadecahedrons (Fig. 2b). There are two types of H atoms labeled as H1 and H2 (occupying two different $2c$ Wyckoff sites): the former is shared by four Te atoms, and the latter, while coordinated by three Te atoms, bonds with another H2 atom, forming also a quasi-molecular $H_2$-unit with a bond length of ~0.86 Å (at 300 GPa). We found a rather weak pressure dependence of the intramolecular H-H bond length for both $P6/mmm$ and $R-3m$ structures, similar to those reported in other compressed hydrides [17, 21, 23, 33].

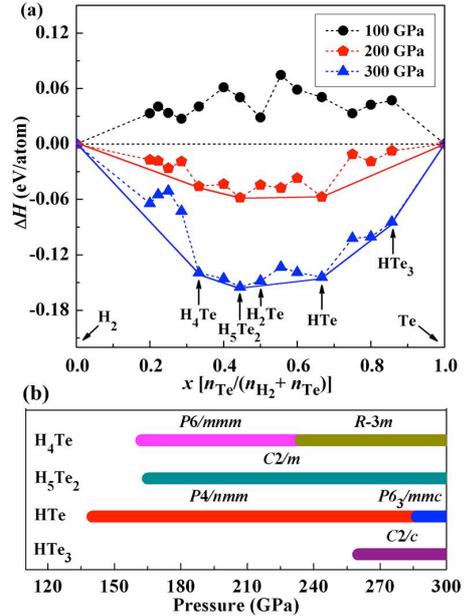

**Fig. 1 (color online).** (a) Formation enthalpies ($\Delta H$) of various H-Te compounds with respect to decomposition into constituent elemental solids at 100-300 GPa. Data points located on the convex hull (solid lines) represent stable species against any type of decomposition. (b) Pressure ranges in which the corresponding structures of different stoichiometries are stabilized. The phases I ($P3_121$), V ($Im-3m$), VI ($I4/mmm$) and VII ($Fm-3m$) of Te [34], and the $P6_3m$ and $C2/c$ structures of solid $H_2$ [35] are used for calculating $\Delta H$ (Table S1). The hull data remain essentially unchanged with the inclusion of zero-point energies (Fig. S2).

The energetically favored structure of $H_5Te_2$ (stable above 165 GPa) has a $C2/m$ symmetry (Fig. 2c), consisting of puckered layers made up of three-fold coordinated Te atoms (at the $4i$ sites) surrounded by two inequivalent H atoms (both at the $4i$ sites, labeled as H1 and H3). Another type of H atoms (at the $2c$ sites, labeled as H2) is accommodated in between two puckered layers and connects with two H1 atoms, forming unexpected linear quasi-molecular $H_3$ units. The two H-H covalent bonds have length of ~0.92 Å with a calculated maximal bonding ELF at ~0.8 (Fig. S3b) at 200 GPa. The formation of $H_3$ units accompanies with a similar charge transfer from Te to H (Table S3). Note that such linear $H_3$ units were often found in gas molecules [36-39], but rarely observed in solid phases (*e.g.* $BaH_6$ [40] and H-rich Rb compounds [41]).

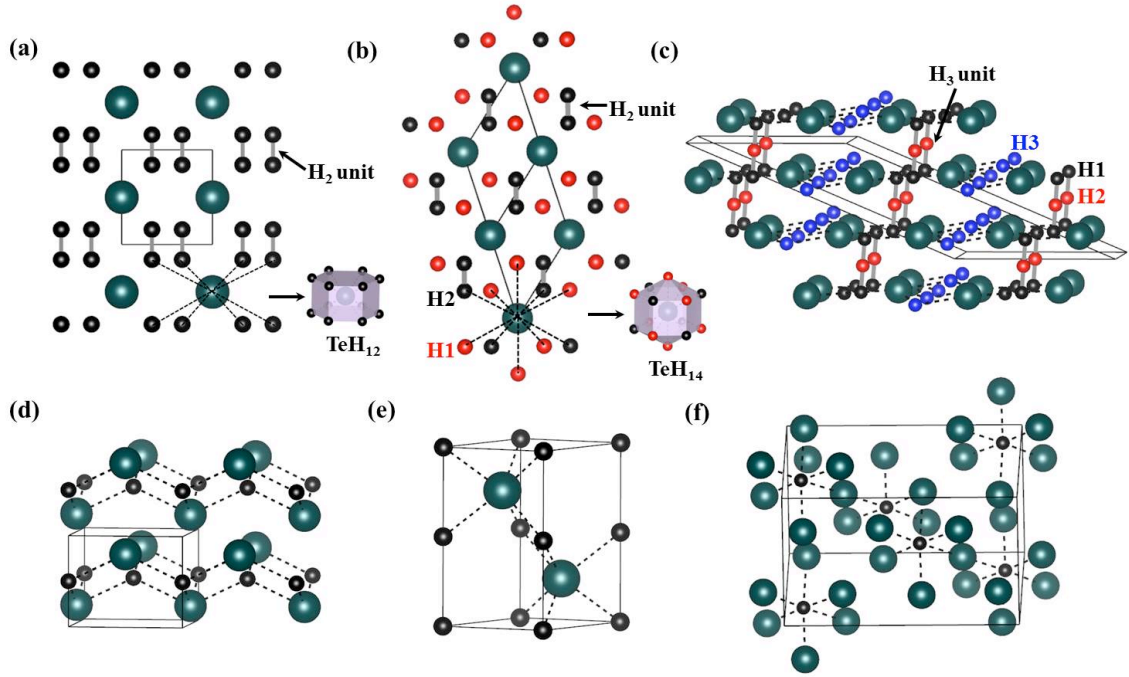

**Fig. 2 (color online).** Structures of stable H-Te compounds at high pressures derived from the CALYPSO structure search: (a, b) $H_4Te$ in the $P6/mmm$ and $R-3m$ structures, respectively; (c) $H_5Te_2$ in the $C2/m$ structure; (d, e) HTe in the $P4/nmm$ and $P6_3/mmc$ structures, respectively; (f) $HTe_3$ in the $C2/c$ structure. Small and large spheres represent H and Te atoms. Solid lines depict the unit cells of the structures. See Table S2 for more detailed structural information.

Turning to the HTe stoichiometry (stable above 140 GPa), it adopts a PbO-type structure (space group $P4/nmm$, Fig. 2d) composed of edge-sharing $TeH_4$-tetrahedron layers, isostructural to that of Fe-based superconductor FeSe [42] and Se hydride of HSe [10]. With increasing pressures, a NiAs-type structure (space group $P6_3/mmc$, Fig. 2e) was stabilized at 286 GPa with an increased coordination number from 4 to 6. For the H-poor $HTe_3$ stoichiometry (stable above 260 GPa), the stable structure has a $C2/c$ symmetry (Fig. 2f), in which two inequivalent Te atoms (at the $4c$ and $8f$ sites) and H atoms (at the $4e$ sites) form interpenetrating polymeric networks. Each H atom is six-fold coordinated by Te.

The electronic band structure and density of states (DOS) of all above H-Te structures (Fig. 3 and Fig. S4) exhibit metallic features. To probe the role of Te played in the electronic structures of H-rich species (*e.g.* $H_4Te$), we calculated the band structure of a hypothetical $H_4Te_0$ system by removing Te out of the lattice, where H sublattice remains unchanged. A uniform compensated background charge (6e/Te) is applied to preserve the total valence electrons of the system. The resultant band structure (red dash lines in Fig. 3a) is essentially similar to the realistic one of $H_4Te$. The indication is that the Te atoms mainly act as electron donors, consistent with the ionic bonding nature of H-Te bonds as described above. This is in contract to the S and Se hydrides where the strong covalent H-S/Se bonding dominates [4,10]. For both $H_4Te$ (Fig. 3a) and $H_5Te_2$ (Fig. 3b), we see a strong DOS peak originating from Te-$p$ and substantial H-derived states around the Fermi level ($E_f$). Meanwhile, there appears "flat band-steep band" characteristic [43] around the $E_f$. These are typical features favorable for strong EPC and thus high-$T_c$ superconductors.

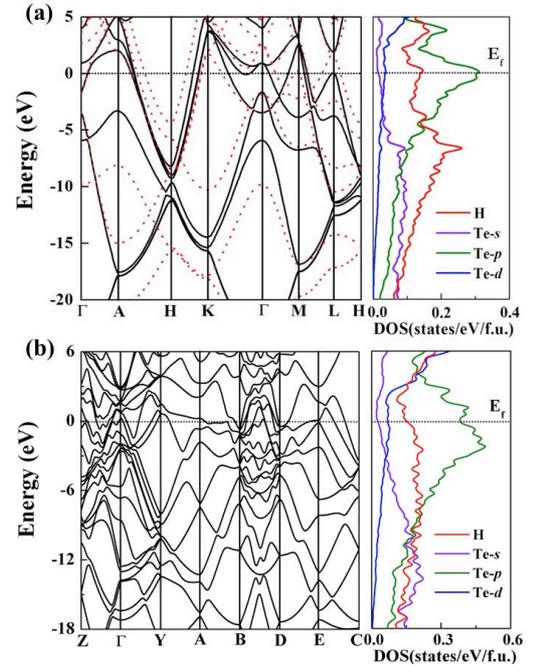

**Fig. 3 (color online).** Electronic band structure and projected density of states for (a) $H_4Te$ in the $P6/mmm$ structure and (b) $H_5Te_2$ in the $C2/m$ structure at 200 GPa. In (a), the red dash lines represent the band structure of the H sublattice with a uniform compensated background charge.

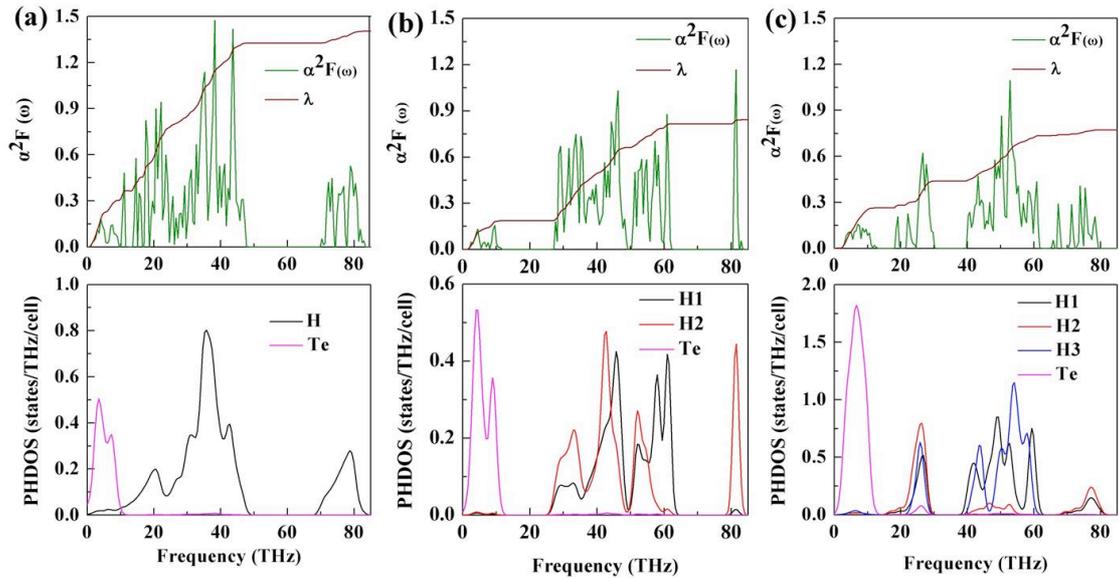

**Fig. 4 (color online).** Calculated projected phonon density of states (lower panels), Eliashberg EPC spectral function $\alpha^2F(\omega)$ and its integral $\lambda(\omega)$ (upper panels) of (a) $H_4Te$ ($P6/mmm$) at 200 GPa, (b) $H_4Te$ ($R\text{-}3m$) at 300 GPa, and (c) $H_5Te_2$ ($C2/m$) at 300 GPa.

We then performed phonon and EPC calculations for three H-rich species of $H_4Te$, $H_5Te_2$ and HTe to probe their potential superconductivity. The absence of any imaginary frequency in their phonon spectra clearly indicates dynamical stability of all the predicted H-Te structures (Fig. S5). The calculated EPC parameter $\lambda$ for $H_4Te$, $H_5Te_2$ and HTe (at 200 GPa) are 1.46, 1.14, and 0.58, respectively. The former two are comparable to the predicted values in $H_2S$ [2] and $H_3S/H_3Se$ [4,10], indicating the fairly strong EPC in Te hydrides. Interestingly, $\lambda$ almost reduces linearly with the decreased H contents in the sequential order of $H_4Te$, $H_5Te_2$, and HTe (Fig. S6). This underlies the dominant role of H played in controlling the EPC of H-rich hydrides [32].

Fig. 4 shows the projected phonon DOS, Eliashberg EPC spectral function $\alpha^2F(\omega)$ and its integral $\lambda(\omega)$ for $H_4Te$ and $H_5Te_2$. For the $P6/mmm$-$H_4Te$ (Fig. 4a), the low-frequency translational vibrations from heavy Te atoms (below 10 THz) contribute to 22% of the total $\lambda$. Regrettably, the high-frequency stretching modes associated with the quasi-molecular $H_2$ units (70-80 THz) contribute to only 5% of $\lambda$. The main contributor to the EPC (72% of $\lambda$) originates from the mid-lying H-Te wagging and H-Te-H bending vibrations (10-50 THz). In the same manner, for the $R\text{-}3m$-$H_4Te$ (Fig. 4b), the high-frequency vibrations from $H_2$ units, intermediate-lying H-derived wagging and bending modes contribute to 3% and 73% of the total $\lambda$, respectively. The similar trend applies to $H_5Te_2$ (Fig. 4c), where $\lambda$ mainly originates from the contribution of mid-lying H-derived phonons (15-60 THz, 62%), with a quite small contribution (<5%) from the high-frequency vibrations of the linear $H_3$ units (70-80 THz). These results highlight that intermediate-frequency H-derived phonons and low-frequency vibrations from Te atoms are mainly responsible for the EPC in H-Te systems. The involvement of massive Te atoms softens these phonon modes, from which the strong EPCs originate. This superconductive mechanism is not unusual by seeing those in $SnH_4$ [21,33] and $CaH_6$ [19] that contain similar quasi-molecular $H_2/H_4$-units. However, it is apparently contrasted to those in H-S and H-Se systems, where the high-frequency H-stretching vibrations dominate the overall EPC [2,4,10].

The superconducting $T_c$ of the predicted H-rich compounds at varied pressures is evaluated through the Allen-Dynes modified McMillan equation [44] by using the calculated logarithmic average frequency $\omega_{log}$ and a typical choice of Coulomb pseudopotential $\mu^* = 0.1$ (as summarized in Table S4). The resultant $T_c$ values are 99 K, 58 K and 19 K for $H_4Te$, $H_5Te_2$ and HTe at 200 GPa, respectively. The high $T_c$ of $H_4Te$ is attributed to the strong EPC ($\lambda = 1.46$) and a reasonably high $\omega_{log}$ (929.1 K). For $H_4Te$ and $H_5Te_2$, we found a negative pressure dependence of $T_c$, which is resulted from the decreased $\lambda$ under compression, even though $\omega_{log}$ increases due to phonon hardening. In the established stable pressure regions for all the compounds (Fig. 1b), the maximum $T_c$ of 104 K occurs in $H_4Te$ at 170 GPa.

Despite of the isoelectronic nature of Te to S/Se, we found via unbiased structure searches an entirely different potential energy landscape of Te hydrides. This sharp distinction originates from fundamentally different chemical bonding behaviors of H-Te and H-S/Se. In S/Se hydrides, S and Se have smaller atomic cores and thus are favorable for forming strong covalent bonds with H. Since electronegativities of S (2.58 by Pauling scale) and Se (2.55) are higher than that of H (2.2), the bonds are polarized towards S/Se (being anion-like). There is no covalent bonding between cation-like H atoms other than weak hydrogen bonds. As for Te hydrides, however, our calculations revealed that the formation of H-Te covalent bonds is energetically unfavorable (evidenced by the strong instability of $TeH_2$ at ambient pressure). Owing to the weaker electronegativity of Te (2.1) than that of H, there exists a notable charge transfer from Te to H. The compressed Te hydrides are stabilized by the formation of ionic H-Te bonds. Upon appearance of the additional charges on H atoms, substantial energy is reduced via the formation of H-H covalent bonds. This might be the physical mechanism on the stabilization of quasi-molecular "$H_2/H_3$" units in the H-rich compounds. The existence of quasi-molecular units facilitates accommodating more H, and this explains the emergence of the highest known H-content of $H_4Te$ in chalcogen hydrides.

The two H-rich hydrides of $H_4Te$ and $H_5Te_2$ are promising superconductors with the highest $T_c$ up to 104 K and 58 K, respectively. Different from the superconductors of S/Se hydrides, current results unraveled an intriguing superconductive mechanism where the intermediate-frequency H-derived vibrations, and low-frequency phonons from massive Te atoms contribute mostly to the EPC. Our finding will inevitably stimulate future experimental study on synthesis of Te hydrides and explore their high-$T_c$ superconductivity at high pressures.

The authors acknowledge the funding supports from China 973 Program under Grant No. 2011CB808200, National Natural Science Foundation of China under Grant No. 11274136, 2012 Changjiang Scholar of Ministry of Education and the Postdoctoral Science Foundation of China under grant 2013M541283. L.Z. acknowledges funding support from the Recruitment Program of Global Experts (the Thousand Young Talents Plan).

*Author to whom correspondence should be addressed: yanggc468@nenu.edu.cn, lijun_zhang@jlu.edu.cn or mym@jlu.edu.cn.

**References**

[1] A. P. Drozdov, M. I. Eremets and I. A. Troyan, arXiv:1412.0460 (2014).
[2] Y. W. Li *et al*., J. Chem. Phys. **140**, 174712 (2014).
[3] E. Cartlidge, Nature, doi:10.1038/nature.2014.16552 (2014).
[4] D. Duan *et al.*, Scientific reports **4**, 6968 (2014).
[5] N. Bernstein *et al.*, arXiv:1501.00196 (2014).
[6] D. Duan *et al.*, arXiv:1501.01784 (2015).
[7] D. A. Papaconstantopoulos *et al.*, arXiv:1501.03950 (2015).
[8] A. P. Durajski, R. Szczesniak and Y. Li, arXiv:1412.8640 (2015).
[9] I. Errea *et al.*, arXiv:1502.02832 (2015).
[10] S. T. Zhang *et al*., arXiv:1502.02607 (2015).
[11] J. A. Flores-Livas, A. Sanna and E. K. U. Gross, arXiv:1501.06336 (2015).
[12] M. Collins, C. Ratcliffe and J. Ripmeester, J. Phys. Chem. **93**, 7495 (1989).
[13] J. H. Loehlin, P. G. Mennitt and J. S. Waugh, J. Chem. Phys. **44**, 3912 (1966).
[14] D. R. Lide, CRC Handbook of Chemistry and Physics (87th ed.). Boca Raton, FL: CRC Press., ISBN 0-8493-0487-8493 (2006).
[15] G. Y. Gao *et al.*, J. Phys. Chem. C **116**, 1995 (2012).
[16] D. Y. Kim *et al.*, Phys. Rev. Lett. **107**, 117002 (2011).
[17] E. Zureka *et al.*, Proc. Natl. Acad. Sci. U.S.A. **106**, 17640 (2009).
[18] P. Baettig and E. Zurek, Phys. Rev. Lett. **106**, 237002 (2011).
[19] H. Wang *et al*., Proc. Natl. Acad. Sci. U. S. A. **109**, 6463 (2011).
[20] L. R. Testardi. Phys. Rev. B **5**, 4342 (1972).
[21] J. S. Tse, Y. Yao and K. Tanaka, Phys. Rev. Lett. **98**, 117004 (2007).
[22] M. I. Eremets *et al.*, Science **319**, 1506 (2008).
[23] G. Gao *et al*., Phys. Rev. Lett. **101**, 107002 (2008).
[24] Y. Wang et al., Phys. Rev. B **82**, 094116 (2010).
[25] Y. Wang, *et al*., Comput. Phys. Commun. **183**, 2063 (2012).
[26] Y. Wang *et al.*, Nat. Commun. **2**, 563 (2011).
[27] L. Zhu *et al*., Nat. Chem. **6**, 645 (2014).
[28] J. Lv *et al.*, Phys. Rev. Lett. **106**, 015503 (2011).
[29] G. Kresse and J. Furthmuller, Phys. Rev. B **54**, 11169. (1996).
[30] J. P. Perdew *et al.*, Phys. Rev. B **46**, 6671 (1992).
[31] P. Giannozzi *et al.*, J. Phys-Condens. Mat. **21**, 395502 (2009).
[32] N. W. Ashcroft, Phys. Rev. Lett. **92**, 187002 (2004).
[33] G. Gao, *et al.*, Proc. Natl. Acad. Sci. U.S.A. **107**, 1317 (2010).
[34] T. Sugimoto, *et al.*, Journal of Physics: Conference Series **500**, 192018 (2014).
[35] C. J. Pickard and R. J. Needs, Nat. Phys. **3**, 473 (2007).
[36] R. Golser *et al*., Phys. Rev. Lett. **94**, 223003 (2005).
[37] H. Gnaser and R. Golser, Phys. Rev. A **73**, 021202 (2006).
[38] W. Wang *et al*., Chem. Phys. Lett. **377**, 512 (2003).
[39] Z. Wang *et al.*, Chem. Sci. **6**, 522 (2015).
[40] J. Hooper *et al.*, J. Phys. Chem. C **117**, 2982 (2013).
[41] J. Hooper and E. Zurek, Chem. Eur. J. **18**, 5013 (2012).
[42] F. C. Hsu, *et al.*, Proc. Natl. Acad. Sci. U.S.A. **105**, 14262 (2008).
[43] A. Simon, Angew. Chem. Int. Ed. **36**, 1788 (1997).
[44] P. B. Allen and R. C. Dynes, Phys. Rev. B 12, 905 (1975).

# Supplementary Information for the paper entitled "*Tellurium Hydrides at High Pressures: High-temperature Superconductors*"

Xin Zhong[1], Hui Wang[1], Jurong Zhang[1], Hanyu Liu[1], Shoutao Zhang[1], Hai-Feng Song[4,5], Guochun Yang[2,1], Lijun Zhang[3,1], Yanming Ma[1]

[1]*State Key Laboratory of Superhard Materials, Jilin University, Changchun 130012, China.*
[2]*Faculty of Chemistry, Northeast Normal University, Changchun 130024, China.*
[3]*College of Materials Science and Engineering and Key Laboratory of Automobile Materials of MOE, Jilin University, Changchun 130012, China*
[4]*LCP, Institute of Applied Physics and Computational Mathematics, Beijing 100088, China*
[5]*Software Center for High Performance Numerical Simulation, China Academy of Engineering Physics, Beijing 100088, China*

# Supplementary Methods

Our structural prediction approach is based on a global minimization of free energy surfaces of given compounds by combining *ab initio* total-energy calculations with the particle swarm optimization (PSO) algorithm.[1,2] The structure search of each $H_xTe_y$ ($x$ = 1−8 and $y$ = 1−3) stoichiometry is performed with simulation cells containing 1−4 formula units. In the first generation, a population of structures belonging to certain space group symmetries are randomly constructed. Local optimizations of candidate structures are done by using the conjugate gradients method through the VASP code[3], with an economy set of input parameters and an energy convergence threshold of 1 × 10$^{-5}$ eV per cell. Starting from the second generation, 60% structures in the previous generation with the lower enthalpies are selected to produce the structures of next generation by the PSO operators. The 40% structures in the new generation are randomly generated. A structure fingerprinting technique of bond characterization matrix is employed to evaluate each newly generated structure, so that identical structures are strictly forbidden. These procedures significantly enhance the diversity of sampled structures during the evolution, which is crucial in driving the search into the global minimum. For most of cases, the structure search for each chemical composition converges (evidenced by no structure with the lower energy emerging) after 1000 ~ 1200 structures investigated (*i.e.* in about 20 ~ 30 generations).

The energetic stabilities of different $H_xTe_y$ stoichiometries are evaluated by their formation enthalpies relative to the products of dissociation into constituent elements (*i.e.* solidified phases of $H_2$ and Te):

$$\Delta H = [h(H_xTe_y) - xh(H) - yh(Te)]/(x+y) \qquad (1)$$

where *h* represents absolute enthalpy. By regarding H and Te as the binary variables, with these $\Delta H$ values we can construct the convex hull at each pressure (Fig. 1a in the main text). It is known that the zero-point energy plays an important role in determining the phase stabilities of the compounds containing light elements such as

H. We hence examine the effect of zero-point energy on the stability of the stoichiometries on the convex hull (as shown in Fig. S2), by using the calculated phonon spectrum with the supercell approach[6] as implemented in the Phonopy code[7].

The electron-phonon coupling calculations are carried out with the density functional perturbation (linear response) theory as implemented in the QUANTUM ESPRESSO package.[8] We employ the norm-conserving pseudopotentials with the $1s^1$ and $5s^2 5p^4$ configurations as valence electrons for H and Te. The kinetic energy cutoff for wave-function expansion is chosen as 80 Ry. To reliably calculate electron-phonon coupling in metallic systems, we need to sample dense $k$-meshes for electronic Brillouin zone integration and enough $q$-meshes for evaluating average contributions from all the phonon modes. Dependent on specific structures of stable compounds, different $k$-meshes and $q$-meshes are used: 16 x 16 x 16 $k$-meshes and 4 x 4 x 4 $q$-meshes for $H_4$Te in the *P6/mmm* structure, 20 x 20 x 20 $k$-meshes and 5 x 5 x 5 $q$-meshes for $H_4$Te in the *R-3m* structure, 20 x 20 x 8 $k$-meshes and 5 x 5 x 2 $q$-meshes for $H_5Te_2$ in the *C2/m* structure, 16 x 16 x 20 $k$-meshes and 4 x 4 x 5 $q$-meshes for HTe with the *P4/nmm* structure, 20 x 20 x 16 $k$-meshes and 5 x 5 x 4 $q$-meshes for HTe with the *P6$_3$/mmc* structure.

The reliabilities of the projected-augmented-wave (PAW) pseudopotentials for H and Te at high pressures, which are used for all the energetic calcualtions, are crosschecked with the full-potential linearized augmented plane-wave (LAPW) method as implemented in the WIEN2k code[9]. By using two different methods, we calculate total energies of $H_4$Te in the *P6/mmm* structure at varied pressures, and then fit the obtained energy-volume data into the Birch-Murnaghan equation of states. Fig. S0 shows the resulted fitted equation of states. We can see the results derived from two methods are almost identical. This clearly indicates the suitability of the PAW pseudopotentials for describing the energetic stabilities of Te hydrides at megabar pressures.

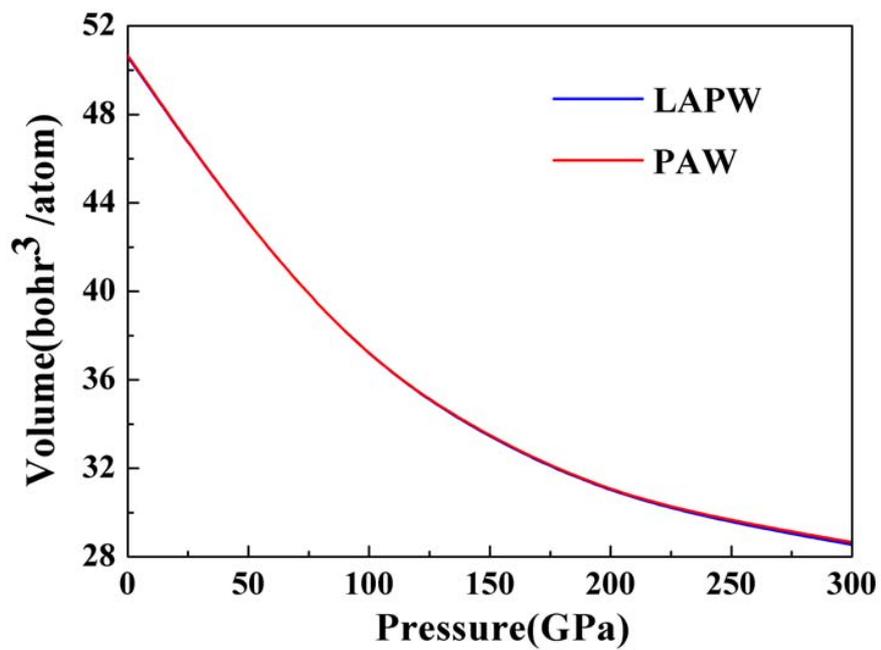

**Figure S0.** Comparison of the fitted Birch-Murnaghan equation of states for $H_4Te$ in the *P6/mmm* structure by using the calculated results from the PAW pseudopotentials and the full-potential LAPW method.

# Supplementary Figures

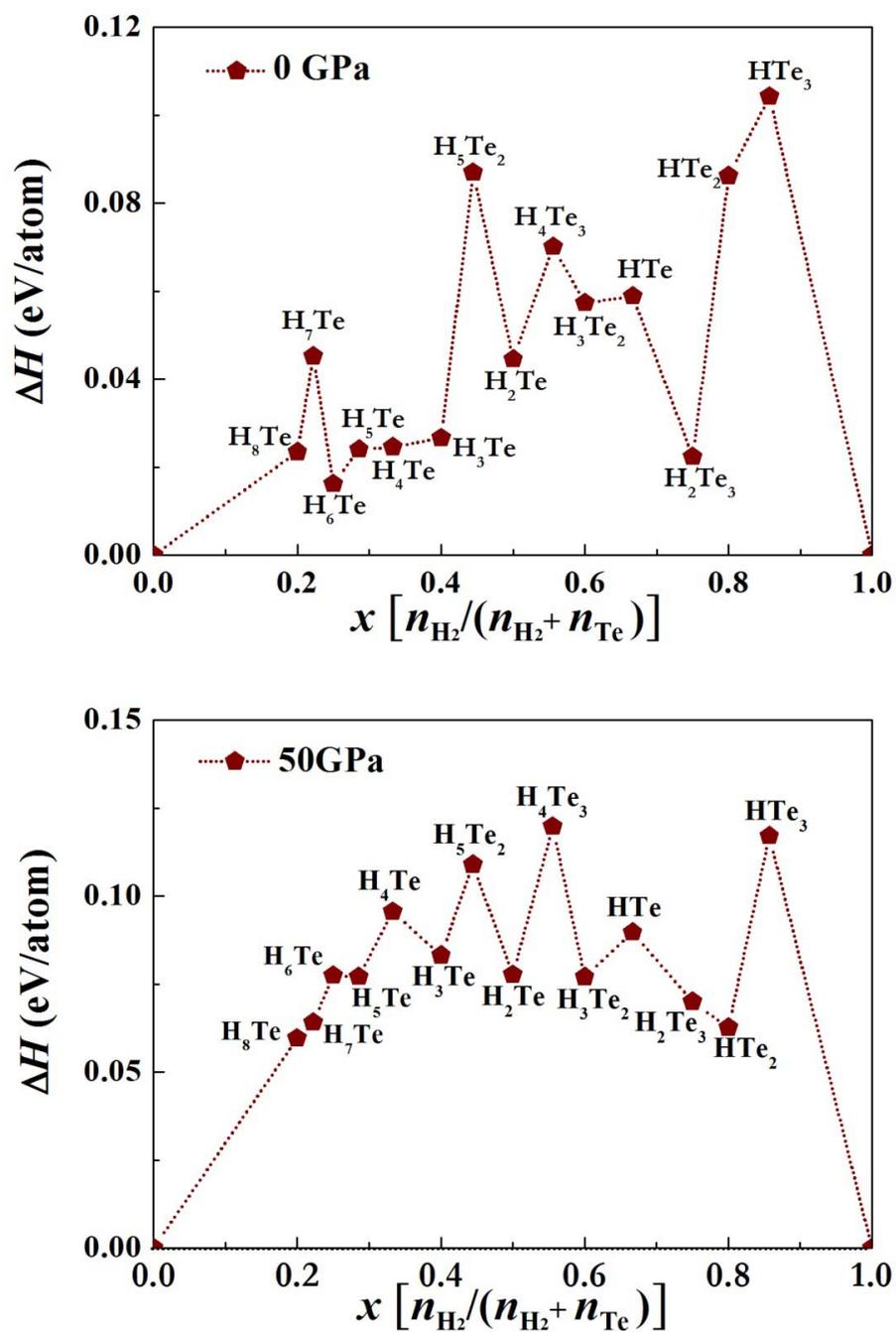

**Figure S1.** The phase stabilities (represented by the convex hulls) of various H-Te compounds at the low pressure range of 0 (upper panel) and 50 GPa (lower panel).

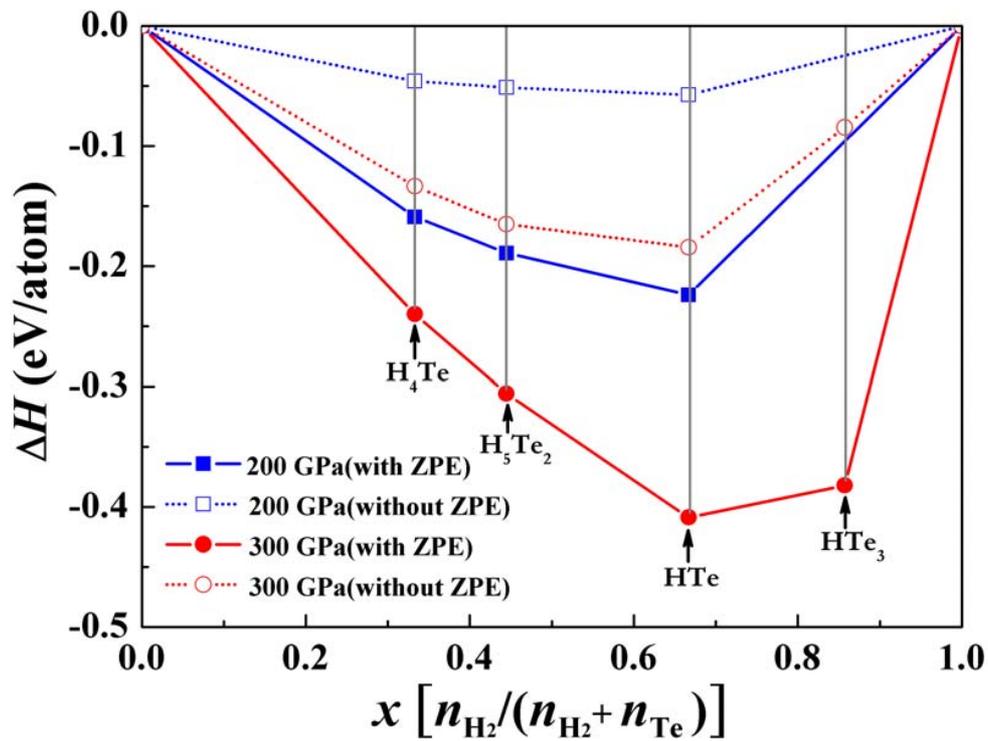

**Figure S2.** Relative formation enthalpies of the uncovered stable stoichiometries, *e.g.* $H_4Te$, $H_5Te_2$, $HTe$ and $HTe_3$ at 200 and 300 GPa with (filled symbols and solid lines) and without (open symbols and dot lines) the inclusion of zero-point energy contribution. One clearly sees that the introduction of zero-point energy does not change the essential energetic stabilities of these stable stoichiometries.

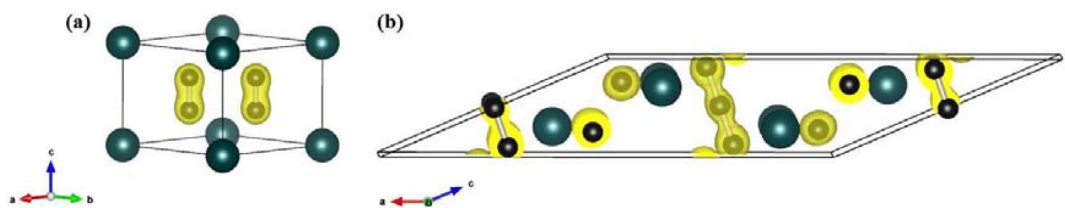

**Figure S3.** Three-dimensional plots of the electron localization function (ELF) for H$_4$Te [$P6/mmm$] (a) and H$_5$Te$_2$ [$C2/m$] (b) at 200 GPa (isosurface value = 0.8).

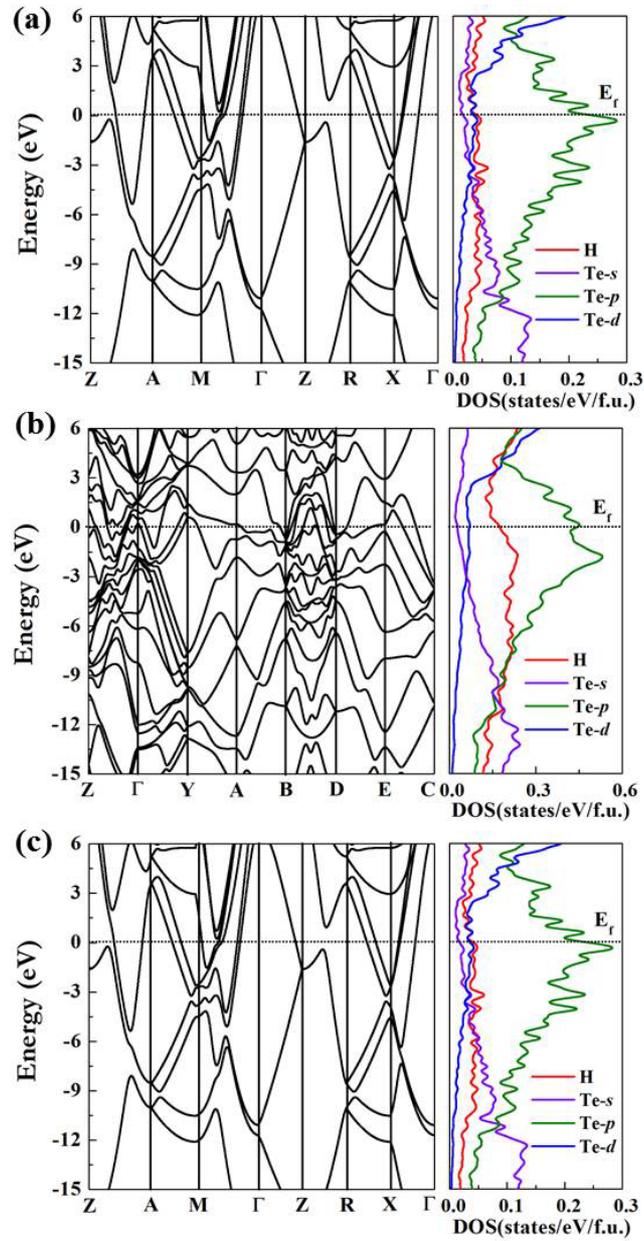

**Figure S4.** Calculated electronic band structure (left panels) and projected density of states (right panels) for the H-rich compounds at the lower boundary of respective stable pressure region: (a) $H_4Te$ [*P6/mmm*] at 162 GPa, (b) $H_5Te_2$ [*C2/m*] at 165 GPa, and (c) HTe [*P4/nmm*] at 141 GPa.

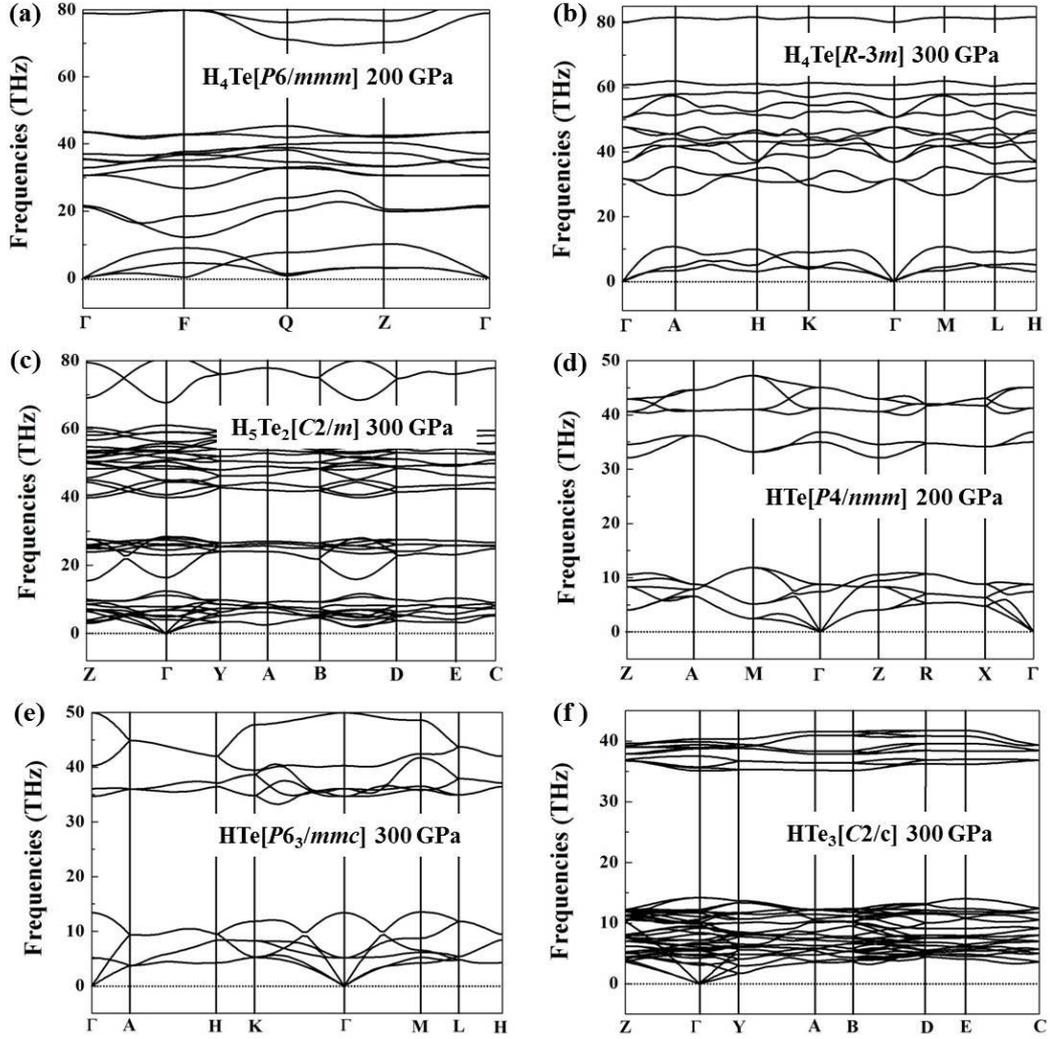

**Figure S5.** Calculated phonon dispersion curves for (a) H$_4$Te [*P*6/*mmm*] at 200 GPa, (b) H$_4$Te [*R*-3*m*] at 300 GPa, (c) H$_5$Te$_2$ [*C*2/*m*] at 300 GPa, (d) HTe [*P*4/*nmm*] at 200 GPa, (e) HTe [*P*6$_3$/*mmc*] at 300 GPa and (f) HTe$_3$ [*C*2/*c*] at 300 GPa. These results unambiguously demonstrate lattice dynamical stabilities of the uncovered stable compounds in view of the absence of any imaginary phonon mode.

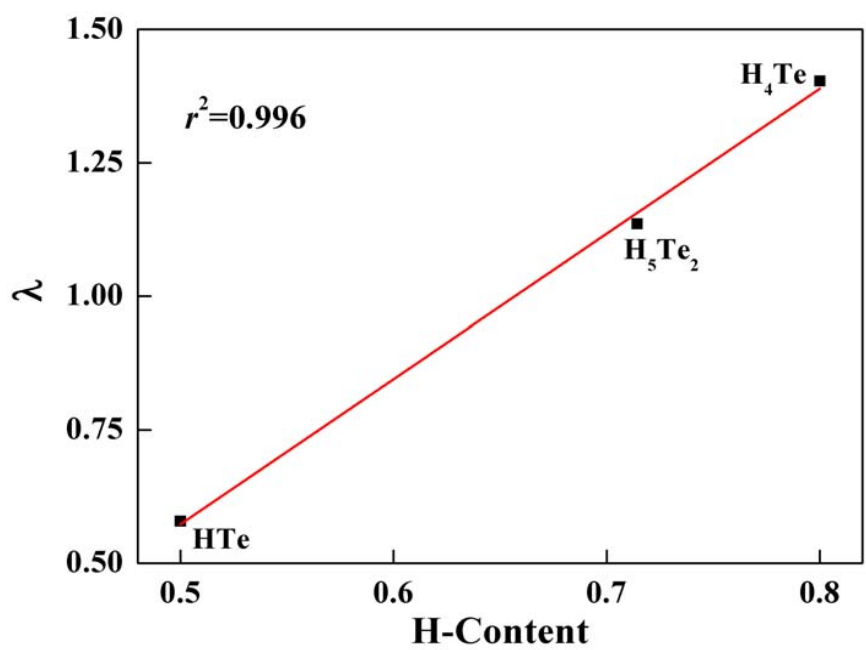

**Figure S6.** Dependence of the total electron-phonon coupling parameter λ on the H content of three H-rich compounds at 200 GPa. The red line represents a linear fitting and r² is the linear correlation coefficient.

# Supplementary Tables

**Table S1.** Detailed structural information of solidified Te and $H_2$ phases used for calculating the decomposition enthalpy.

| Phase | Pressure (GPa) | Lattice parameters (Å) | Atomic coordinates (fractional) | | | |
|---|---|---|---|---|---|---|
| Te I ($P3_121$) | 0 | $a = b = 4.5084$<br>$c = 5.957$<br>$\alpha = \beta = 90.000$<br>$\gamma = 120.000$ | Te(1a) | 0.000 | 0.730 | 0.167 |
| Te V ($Im\text{-}3m$) | 50 | $a = b = c = 3.3961$<br>$\alpha = \beta = \gamma = 90.000$ | Te(1a) | 0.500 | 0.500 | 0.500 |
| Te VI ($I4/mmm$)[a] | 100 | $a = b = 2.8553$<br>$c = 4.0789$<br>$\alpha = \beta = \gamma = 90.000$ | Te(1a) | 0.500 | 0.500 | 0.000 |
| Te VII ($Fm\text{-}3m$) | 200 | $a = b = c = 3.8226$<br>$\alpha = \beta = \gamma = 90.000$ | Te(1a) | 0.000 | 0.000 | 0.500 |
| H-$P6_3m$ | 50 | $a = b = 5.1680$<br>$c = 4.1700$<br>$\alpha = \beta = 90.000$<br>$\gamma = 120.000$ | H(1a) | 0.333 | 0.667 | 0.840 |
| H-$C2/c$ | 200 | $a = 3.000$<br>$b = 5.186$<br>$c = 6.115$<br>$\alpha = \gamma = 90$<br>$\beta = 119.50$ | H(8f) | 0.660 | 0.224 | 0.626 |

[a]The structure of the Te VI phase is not experimentally confirmed yet. Here, we take the lowest-enthalpy $I4/mmm$ structure from the structure search we carried out.

**Table S2.** Detailed structural information of the uncovered stable H-Te compounds at selected pressures.

| Phase | Pressure (GPa) | Lattice parameters (Å) | Atomic coordinates (fractional) | | | |
|---|---|---|---|---|---|---|
| HTe$_3$-*C2/c* | 300 | $a$ = 4.514<br>$b$ = 7.985<br>$c$ = 4.505<br>$\alpha = \gamma$ = 90.000<br>$\beta$ = 71.804 | H(4e)<br>Te(8f)<br>Te(4c) | 0.500<br>0.240<br>0.250 | 0.576<br>0.586<br>0.750 | 0.250<br>0.994<br>0.500 |
| HTe-*P4/nmm* | 200 | $a = b$ = 3.376<br>$c$ = 2.705<br>$\alpha = \beta = \gamma$ = 90.000 | H(2b)<br>Te(2c) | 0.500<br>0.500 | 0.500<br>0.000 | 0.500<br>0.801 |
| HTe-*P6$_3$/mmc* | 300 | $a = b$ = 2.621<br>$c$ = 4.549<br>$\alpha = \beta$ = 90.000 | H(2a)<br>Te(2d) | 0.000<br>0.333 | 0.000<br>0.667 | 0.000<br>0.750 |
| H$_5$Te$_2$-*C2/m* | 200 | $a$ = 10.092<br>$b$ = 3.186<br>$c$ = 5.550<br>$\alpha = \gamma$ = 90.000<br>$\beta$ = 156.680 | H1(4i)<br>H2(2c)<br>H3(4i)<br>Te(4i) | 0.230<br>0.000<br>0.160<br>0.738 | 0.000<br>0.000<br>0.500<br>0.000 | 0.897<br>0.500<br>0.257<br>0.714 |
| H$_4$Te-*P6/mmm* | 200 | $a = b$ = 2.944<br>$c$ = 2.686<br>$\alpha = \beta$ = 90.000<br>$\gamma$ = 120.000 | H(4h)<br>Te(1a) | 0.333<br>0.000 | 0.667<br>0.000 | 0.659<br>0.000 |
| H$_4$Te-*R-3m* | 300 | $a = b = c$ = 3.051<br>$\alpha = \beta = \gamma$ = 54.712 | H(2c)<br>H(2c)<br>Te(a) | 0.555<br>0.222<br>0.000 | 0.555<br>0.222<br>0.000 | 0.555<br>0.222<br>0.000 |

**Table S3.** Residual charges on H and Te atoms based on the Bader charge analysis in H$_4$Te [*P6/mmm*] and H$_5$Te$_2$ [*C2/m*] at 200 GPa. σ represents the charge transferred from Te to H.

|  | Atom | Charge | σ(*e*) |
|---|---|---|---|
| **H$_4$Te** | Te | 5.10 | 0.90 |
|  | H | 1.22 | -0.22 |
| **H$_5$Te$_2$** | Te | 5.37 | 0.63 |
|  | H1 | 1.31 | -0.31 |
|  | H2 | 1.14 | -0.11 |
|  | H3 | 1.25 | -0.21 |

**Table S4.** Calculated electron-phonon coupling parameter λ, logarithmic average frequency $\omega_{\log}$ and critical temperature $T_c$ of the uncovered H-rich compounds: H$_4$Te in the *P6/mmm* structure (170, 200 and 230 GPa) and the *R-3m* structure (270 and 300 GPa); H$_5$Te$_2$ (200 and 300 GPa); HTe in the *P4/nmm* structure (150 and 200 GPa) and the *P6$_3$/mmc* structure (300 GPa).

|  | P (GPa) | λ | $\omega_{\log}$ (K) | $T_c$ (K) |
|---|---|---|---|---|
| **H$_4$Te** | 170 | 1.461 | 940.249 | 104.47 |
|  | 200 | 1.403 | 929.110 | 99.18 |
|  | 230 | 1.422 | 844.188 | 91.33 |
|  | 270 | 0.923 | 1237.535 | 75.66 |
|  | 300 | 0.842 | 1307.098 | 67.70 |
| **H$_5$Te$_2$** | 200 | 1.135 | 790.315 | 57.98 |
|  | 300 | 0.772 | 1058.281 | 46.00 |
| **HTe** | 150 | 0.712 | 781.559 | 28.28 |
|  | 200 | 0.579 | 914.586 | 18.71 |
|  | 300 | 1.0723 | 573.634 | 44.26 |

**Reference**

(1) Wang, Y.; Lv, J.; Zhu, L.; Ma, Y. Crystal Structure Prediction via Particle-Swarm Optimization. *Phys. Rev. B* **2010**, *82*, 094116.

(2) Wang, Y.; Lv, J.; Zhu, L.; Ma, Y. CALYPSO: A Method for Crystal Structure Prediction. *Comput. Phys. Commun.* **2012**, *183*, 2063.

(3) Kresse, G.; Furthmuller, J. Efficient Iterative Schemes for ab initio Total-Energy Calculations Using a Plane-Wave Basis Set. *Phys. Rev. B* **1996**, *54*, 11169.

(4) Perdew, J. P.; Chevary, J. A.; Vosko, S. H.; Jackson, K. A.; Pederson, M. R.; Singh, D. J.; Fiolhais, C. Atoms, Molecules, Solids, and Surfaces: Applications of the Generalized Gradient Gpproximation for Exchange and Correlation. *Phys. Rev. B* **1992**, *46*, 6671.

(5) Feng, J.; Hennig, R. G.; Ashcroft, N. W. & Hoffmann, R. Emergent reduction of electronic state dimensionality in dense ordered Li–Be alloys. *Nature*, **2008**, *451*, 445.

(6) Parlinski, K.; Li, Z. & Kawazoe, Y. First-Principles Determination of the Soft Mode in Cubic ZrO2. *Phys. Rev. Lett.* **1997**, *78*, 4063.

(7) Togo, A.; Oba, F. & Tanaka, I. First-principles calculations of the ferroelastic transition between rutile-type and $CaCl_2$-type $SiO_2$ at high pressures. *Phys. Rev. B* **2008**, *78*, 134106.

(8) Giannozzi, P.; Baroni, S.; Bonini, N.; Calandra, M.; Car, R.; Cavazzoni, C.; Ceresoli, D.; Chiarotti, G.; Cococcioni, M.; Dabo, I.; Corso, A.; de Gironcoli, S.; Fabris, S.; Fratesi, G.; Gebauer, R.; Gerstmann, U.; Gougoussis, C.; Kokalj, A.; Lazzeri, M.; Martin-Samos, L.; Marzari, N.; Mauri, F.; Mazzarello, R.; Paolini, S.; Pasquarello, A.; Paulatto, L.; Sbraccia, C.; Scandolo, S.; Sclauzero, G.; Seitsonen, A.; Smogunov, A.; Umari, P.; Wentzcovitch, R. QUANTUM ESPRESSO: a Modular and Open-source Software Project for Quantum Simulations of Materials. *J. Phys.: Condens. Matter.* **2009**, *21*, 395502.

(9) Kunes, J.; Arita, R.; Wissgott, P.; Toschi, A.; Ikeda H.; Held, K. *Comp. Phys. Commun.* **2010**, *181*, 1888–1895.